\documentclass[iop]{emulateapj}

\usepackage{graphicx}
\usepackage{epsfig}
\usepackage{epstopdf}
\newcommand{\CaIIIR}{Ca~II~8542}
\newcommand{\Halpha}{H\ensuremath{\alpha}}
\newcommand{\kms}{km~s$^{-1}$}

\begin{document}

\title{Statistical properties of the disk counterparts of type II spicules from simultaneous observations of RBEs in \CaIIIR\ and \Halpha}

\author{D. H. Sekse\altaffilmark{1}}
\author{L. Rouppe van der Voort\altaffilmark{1}}
\author{B. De Pontieu\altaffilmark{2}}

\affil{\altaffilmark{1}Institute of Theoretical Astrophysics,
  University of Oslo, %
  P.O. Box 1029 Blindern, N-0315 Oslo, Norway}
  
\affil{\altaffilmark{2}Lockheed Martin Solar \& Astrophysics Lab, Org.\ A021S,
  Bldg.\ 252, 3251 Hanover Street Palo Alto, CA~94304 USA}

\begin{abstract}

Spicules were recently found to exist as two different types when a new class of so-called type II spicules was discovered at the solar limb with the Solar Optical Telescope onboard the Hinode spacecraft.  These type II spicules have been linked with on-disk observations of Rapid Blue-shifted Excursions (RBEs) in the \Halpha\ and \CaIIIR\ lines.  Here we analyze observations optimized for the detection of RBEs in both \Halpha\ and \CaIIIR\ lines simultaneously at a high temporal cadence taken with the CRisp Imaging SpectroPolarimeter (CRISP) at the Swedish Solar Telescope (SST) on La Palma.  In this study we used a high-quality time sequence for RBEs at different blue-shifts and employed an automated detection routine to detect a large number of RBEs in order to expand on the statistics of RBEs.  We find that the number of detected RBEs is strongly dependent on the associated Doppler velocity of the images on which the search is performed. Automatic detection of RBEs at lower velocities increases the estimated number of RBEs to the same order of magnitude expected from limb spicules. This shows that RBEs and type II spicules are indeed exponents of the same phenomenon.  Furthermore, we provide solid evidence that \CaIIIR\ RBEs are connected to \Halpha\ RBEs and are located closer to the network regions with the \Halpha\ RBEs being a continuation of the \CaIIIR\ RBEs.  Our results show that RBEs have an average lifetime of 83.9 seconds when observed in both spectral lines and that the Doppler velocities of RBEs range from 10 to 25~\kms\  in \CaIIIR\, and 30 to 50~\kms\ in \Halpha. In addition, we automatically determine the transverse motion of a much larger sample of RBEs than previous studies, and find that just like type II spicules, RBEs undergo significant transverse motions of order 5--10~\kms. Finally, we find that the intergranular jets discovered in Big Bear Solar Observatory are a subset of RBEs.

\end{abstract}

\section{Introduction} 
\label{sec:intro}

Observing the chromosphere at the solar limb, a great number of thin jet-like features called spicules can be seen protruding outwards from the limb.  
While spicules have been known and studied for a long time 
\citep[for an extensive overview, see][]{1968SoPh....3..367B, 
2000SoPh..196...79S}
, progress in understanding their nature has been relatively modest.
However, interest for this phenomenon has revived considerably in recent years. 
The seeing-free environment and aperture of sufficient size for high spatial resolution allowed the {\it Hinode} satellite to provide an unprecedented view of the solar limb in the form of high temporal resolution, long duration time series of Ca~II~H filtergrams. 
From these observations, a new class of spicules was discovered
\citep{2007PASJ...59S.655D} 
that is characterized by short life times (typically 10--100~s), vigorous dynamics, and exclusively upward motion (50--150~\kms). 
Their nature is quite different from what are now considered as classical spicules, or type I spicules, which have life times from 5 to 10 min, and display clear rise and downfall phases at more modest velocities (20--30~\kms).
The characteristic sideways swaying motion that this newly identified class (or type II spicules) displays, has been regarded as a sign that the chromosphere is permeated with Alfv{\'e}nic waves of sufficient energy to accelerate the solar wind and potentially heat the quiet corona  
\citep{2007Sci...318.1574D}. 
Later studies have further strengthened the conjecture that type II spicules have an important role in mass loading and heating of the corona
\citep{2009ApJ...701L...1D, 
2011Sci...331...55D} 

The driving mechanism behind type I spicules seems to be well understood. 
Type I spicules have been identified as the off-limb manifestation of what is known as dynamic fibrils in active regions on the disk. 
The observations and modeling of
\citet{2004Natur.430..536D},
\citet{2006ApJ...647L..73H} 
and \citet{2007ApJ...655..624D} 
demonstrated that dynamic fibrils are driven by magnetoacoustic shocks that result from flows and waves that leak along magnetic field lines from the photosphere into the chromosphere.
A similar mechanism is responsible for driving part of the quiet Sun mottles 
\citep{2007ApJ...660L.169R} 
which helps explain why type I spicules can be detected in both active regions and quiet Sun. 

In contrast to type I spicules, the formation and driver of type II spicules is not well understood and still under debate.
\citet{2011ApJ...736....9M} describe in detail the formation of what resembles a type II spicule in their 3-dimensional radiative magneto-hydrodynamics simulation.
The simulated jets shows similarities to type II spicules in that rapid ejection of cool chromospheric plasma into the corona occurs while it is being heated to temperatures that are high enough to cause apparent fading in typical chromospheric diagnostics.
In the simulation, small-scale flux emergence leads to chromospheric plasma being accelerated by a strong, mostly horizontal, Lorentz force into a region of very strong vertical magnetic field.
The subsequent increase in gas pressure induces the ejection of plasma into the low-density corona while being heated to coronal temperatures by heating processes at the footpoint. 

While this scenario is tantalizing, further modeling is required in order to firmly establish the physics behind type II spicules. 
Perhaps more importantly, high-quality observations are essential in order to constrain the models and guide the direction of investigation.
A complicating factor here is that due to their narrow spatial extents, short lifetimes and significant displacement, as well as the superposition of many spicules along the line of sight, spicules are notoriously difficult to properly measure at the limb.
Finding the disk-counterparts of type II spicules makes it possible to separate them spatially and overcome the problem of superposition encountered at the limb.

\citet{2008ApJ...679L.167L} 
investigated so called "rapid blueshifted excursions" (RBEs) in on-disk \CaIIIR\ data obtained with the Interferometric BIdimensional Spectrometer (IBIS) at the Dunn Solar Telescope.
In their data, events with short-lived blue-ward asymmetries of the \CaIIIR\ line were tentatively linked to type II spicules. 
\citet[][Paper I]{2009ApJ...705..272R} used high-quality \Halpha\ and \CaIIIR\ observations from the Crisp Imaging Spectropolarimeter (CRISP) at the Swedish Solar Telescope (SST) to make a firm connection between these RBEs and type II spicules.
%
They found a large number of RBEs %
for which properties such as lengths, velocities, and life times were found to agree well with what was measured for type II spicules. 
In addition, the appearance, including both swaying motions and apparent propagation away from magnetic field concentrations in the photosphere, was considered as a clear indication that RBEs are the disk counterpart of type II spicules. 

Recently, \citet{2011ApJ...730L...4J} challenged the identification of  RBEs as type II spicules by arguing that the global occurrence rate of RBEs as inferred from the Paper~I measurements is too low.
In this work, we further investigate the occurrence rate of RBEs in new high-quality CRISP observations as well as the observations used in Paper~I. 
For the new observations, we employed a dual-line program that provides both spectrally well sampled \CaIIIR\ and \Halpha\ profiles at relatively high temporal cadence. 
This program allows for a direct comparison between RBEs measured in \CaIIIR\ and \Halpha, a comparison that could only be done indirectly from the sequentially recorded data sets used in Paper~I.
Furthermore, we expand on previous results (from Paper~I) by extending the automated detection routine for RBEs so that we can automatically determine life times and transversal motion for a much larger statistical sample than in Paper~I.

\section{Observations and data reduction}

\begin{figure*}[!t]
\begin{center}
\includegraphics[width=\textwidth]{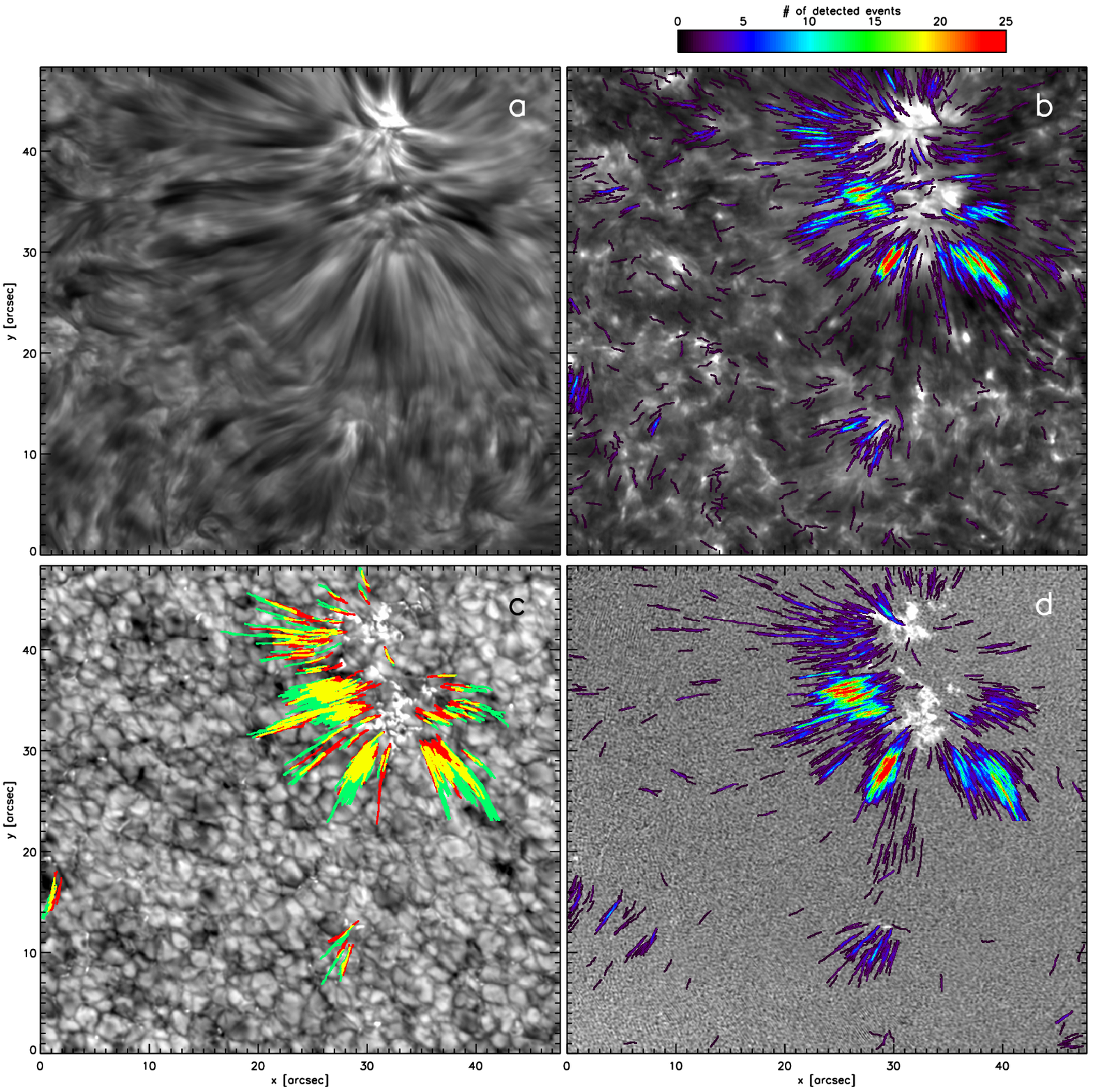}
\caption{Overview of the observed target area. (a): Halpha line core image. (b): \CaIIIR\ line core image with overplotted in rainbow color scale a density map of identified RBEs detected at $-20$~\kms\ over the whole time series. (c): \Halpha\ wideband image (FWHM=4.9\AA\ CRISP prefilter) with overdrawn temporally resolved RBEs that are identified in both \CaIIIR\ and \Halpha: the \CaIIIR\ RBEs are drawn in red, the \Halpha\ RBEs are green, the region where they overlap is yellow. (d): \CaIIIR\ $-600$~m\AA\ Stokes V image in gray color scale with overplotted a density map of the number of identified RBEs detected in \Halpha\ at $-$45~\kms\ during the whole time series. }
\label{fig:noevents}
\end{center}
\end{figure*}

The data was obtained from observations at the Swedish 1-m Solar Telescope 
\citep[SST,][]{2003SPIE.4853..341S} 
on La Palma using the CRisp Imaging SpectroPolarimeter \citep[CRISP,][]{2008ApJ...689L..69S} instrument.  CRISP is equipped with a dual Fabry-P\'{e}rot interferometer system and three high-speed low-noise CCD cameras.  These cameras operate at a frame rate of 35 frames per second with an exposure time of 17 ms per frame and are synchronised by an optical chopper.  Two of the cameras are positioned behind the FPI and a polarizing beam splitter, while the third camera, which is used as an anchor channel for image processing, is positioned before the FPI but after the CRISP pre-filter and is referred to as the wide-band channel.  The field of view (FOV)  with the CRISP instrument is roughly 61$\times$61 arcseconds with an image scale of 
0\farcs0592/pixel. 
CRISP allows for fast wavelength tuning ($<$50 ms) between any two positions within a spectral region given by the spectral width of the prefilter, which makes it ideal for studies of the dynamic time evolution of the chromosphere through imaging spectroscopy.
Here we are interested in the \Halpha\ and \CaIIIR\ spectral lines.
For \Halpha, CRISP has a transmission FWHM of 66~m\AA\ and pre-filter FWHM of 4.9 \AA. For \CaIIIR, the transmission FWHM is 111~m\AA\ and pre-filter FWHM is 9.3\AA.    
High spatial resolution down to the telescope diffraction limit ($\lambda/D$=0\farcs14 and 0\farcs18 for \Halpha\ and \CaIIIR\ respectively) is accomplished by use of the SST adaptive optics system 
\citep[SST,][]{2003SPIE.4853..370S} 
and the Multi-Object, Multi-Frame Blind Deconvolution image restoration technique \citep[MOMFBD][]{2005SoPh..228..191V}.

A 38-minute dataset obtained on June 27, 2010, (17:35-18:13 UT) has been analyzed.
The target area was a coronal hole with a unipolar magnetic island, at solar coordinates (x,y)$\approx$($-$70\arcsec,508\arcsec), ($\mu$=0.84).  
Figure~\ref{fig:noevents} gives an overview of the target region. 

The observational program was made specifically for the detection of RBEs in \CaIIIR\ and \Halpha, simultaneously, with a rather high cadence so that the time evolution of RBEs could be resolved properly. 
The prefilter wheel limited the amount of dead time to 0.63~s when switching between the two lines.  For the \CaIIIR\ line, we ran a program with symmetric sampling at 7 positions in each of the two wings, from  $\pm$120 to $\pm$1452 with steps of 222~m\AA.  In addition, we observed the line core and the blue wing at $-$600~m\AA\ with modulation of the liquid crystals with the purpose of obtaining photospheric magnetograms.  For \Halpha\ we ran a program with dense sampling of 28 positions in the blue wing, from $-$300~m\AA\ to $-$2082~m\AA\ with steps of 66~m\AA.  More detailed information on the sampling positions and their respective Doppler velocities can be seen in Fig. \ref{fig:rbedetail} d and c, respectively.  This program was based on our experiences from Paper I and specifically designed to simultaneously detect RBEs in both lines. RBEs are detected in Ca 8542 in Dopplermaps constructed from the subtraction of symmetric positions from line centre, and in Halpha in maps constructed from the subtraction of the far blue wing image from positions closer to line centre. The program is sufficiently fast (11.8~s cadence) to allow for temporal analysis of RBEs. The symmetric sampling of the Ca 8542 line allows for verification that the detected events are indeed asymmetric in the blue wing which is the defining characteristic of RBEs.

We acquired 8 exposures per line position which were used for Multi-Object, Multi-Frame Blind Deconvolution image restoration \citep[MOMFBD][]{2005SoPh..228..191V}. Exposures from each spectral line were processed in separate MOMFBD restorations. Precise alignment between the wideband and narrowband cameras is achieved by a separate alignment procedure involving a reference pinhole array target. For the MOMFBD restoration of the sequentially recorded CRISP images, the wideband channel serves as a so-called anchor channel that ensures accurate alignment between the different line positions. For certain seeing conditions, the assumption of the size of the iso-planatic patch used to divide the FOV in subimages (here 128 $\times$ 128 pixels, or 7\farcs6 squared) is not sufficiently accurate and the derived PSF is not the optimal solution over the whole area of the subimage. This results in remaining rubber sheet deformations between the CRISP line positions in a single restoration. To reduce this effect, we employed an extra step in the processing following an idea from 
Henriques (private communication). 
%
The wideband images are used twice in a restoration: 1. all images combined in the normal way resulting in a single restored "anchor" image, 2. separated in sets associated with the corresponding CRISP line positions resulting in one restored wideband image per CRISP line position. 
Even though individual wideband exposures enter twice in the restoration procedure,  they are only used once for the determination of the PSF. For the use of the wideband images separated in sets per CRISP line position, the weight is put to zero for the wavefront sensing. The separately restored wideband images have the same rubber sheet deformations as the corresponding CRISP line positions which can now be accurately measured by cross-correlation to the anchor wideband images. The deformation grid determined from the wideband images is then applied to the CRISP images which results in precise alignment between the different line positions. The data analyzed in this study benefited significantly from this extra step in the processing.

After the MOMFBD restoration the images from the individual line scans were combined to form a time series and the \CaIIIR\ images in each scan were aligned with respect to the \Halpha\ images from the same time step.
This alignment between the two lines was done using the wideband images which show photospheric scenes for both spectral lines so that the alignment is done with high accuracy.  The images from both lines were then de-rotated to account for diurnal field rotation, aligned and de-stretched to remove warping due to seeing effects by determining local offsets on the wideband images and applying these to the CRISP images.

For the \CaIIIR\ $-600$ m\AA\ magnetograms, all 32 exposures of the four different liquid crystal states were included in a single MOMFBD restoration, following the scheme described by 
\citet{2008A&A...489..429V} 
and 
\citet{2011A&A...534A..45S}. 
 For the polarimetric response of the telescope at 8542~\AA, we used the model derived by 
 \citet{jaime2010PhDthesis} %
 \citep[also see ][]{2011A&A...527L...8D}. 
 With the 17~ms exposure time, the effective acquisition time for the resulting Stokes V magnetogram is 544~ms.  The noise level is estimated at 0.14\% of the continuum.  It should be noted that at $-600$~m\AA, the Ca 8542 magnetogram is dominated by photospheric signals and is not affected by strong Doppler shifts. Strong Doppler shifts make single wavelength magnetograms closer to the line core, that are more chromospheric in nature, difficult to interpret.  Furthermore, it should be noted that the \CaIIIR\ line is less sensitive to polarization than for example more traditional diagnostics, like the Fe~I~630 lines. In addition, the wavelength position used here, $-600$~m\AA, is relatively less sensitive to polarization than positions closer to the line core.  All these factors combine to a rather limited diagnostic value of the magnetograms used in this study, in particular for the weak magnetic field environment of the coronal hole observed here: the strongest signals barely reach 1\% polarization. 
Despite these limitations, the magnetograms serve as a reliable indicator of the position and polarity of the strongest magnetic field concentrations. These concentrations appear as photospheric bright points in continua and in the far wings of the spectral lines studied here.

We acquired another high-quality data set with the same observing program during the 2010 SST observing campaign: on 03-Jul-2010, we observed a coronal hole at (x,y)$\approx$(19,862) ($\mu$=0.41) for 45 min. In this paper, we present results from the analysis of the 27-Jun-2010 data set but the 03-Jul-2010 data set is used as a reference to verify and check our results and conclusions. 

\section{Method}
\label{sect:method}

\begin{figure}[!ht]
\includegraphics[width=\columnwidth]{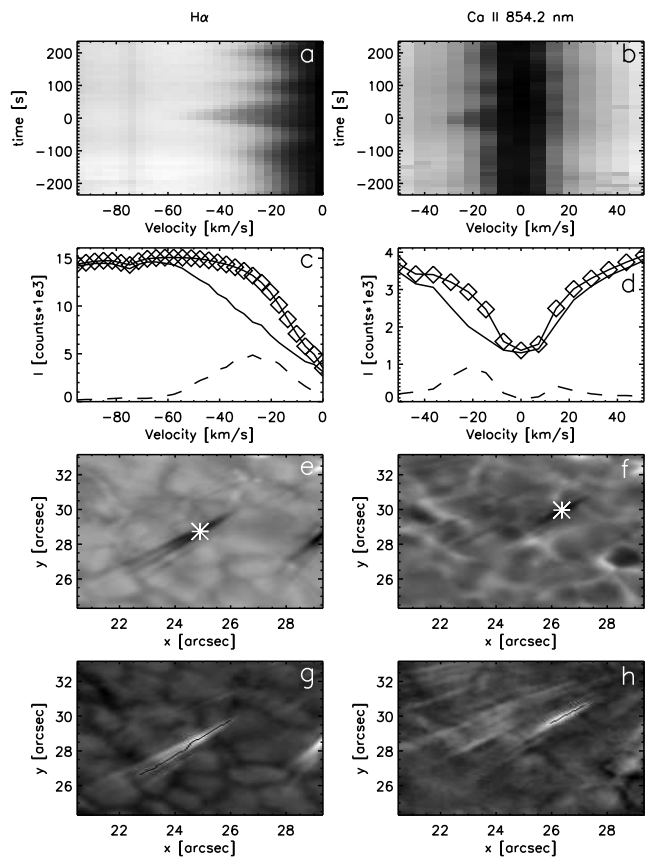}
\caption{Details of an RBE observed in both \Halpha\ (left column) and \CaIIIR\ (right column). Panels (a) and (b) show the temporal evolution of the spectrum at a central location in the RBE, marked with asterices in panels (e) and (f) (note that this is not the same spatial location for the two spectral lines). The temporal axis is centred on the time used for the panels below. Panels (c) and (d) show the spectral profile at the asterix' location with the solid line, the average spectrum over the full FOV with the solid line with diamonds at the sampling points, and the difference between these profiles as the dashed line. Panels (d) and (e) show blue wing images at $-$45~\kms\ for \Halpha\ and $-$20~\kms\ for \CaIIIR. The RBE is seen as a dark streak which was seen to originate from the magnetically enhanced region in the upper right direction in the time steps before. The \CaIIIR\ RBE is shorter and closer to this region than the \Halpha\ RBE. Panels (g) and (h) show the maps used for automated detection: a difference image in \Halpha\ with the $-$45~\kms\ image subtracted from the far wing $-$2~\AA\ image, and a \CaIIIR\ Doppler map at $\pm$20~\kms. The detection skeleton from the automated method is marked with a thin black line. }
\label{fig:rbedetail}
\end{figure}

The dataset was searched for RBEs in the \CaIIIR\ and \Halpha\ lines separately, where they appear in the spectral wings as roundish or elongated dark absorption features against the photospheric background.  
An automated algorithm that locates RBEs by isolating the location and time of a long thin feature at high blueshifts 
was used to find a large number of RBEs in both datasets and both spectral lines in order to get reliable RBE statistics.  
Apart from some minor refinements, the method is identical to the method used for Paper~I.

To make a clean detection of the upper chromospheric RBEs, the background, which often contains linear photospheric features (such as intergranular lanes), has to be reduced or removed from the images.
For \CaIIIR\ this is done by subtraction of the opposite wing position, thereby constructing a Doppler map, and for \Halpha\ this is done by subtracting the far blue wing image, at 2\AA, creating a difference image.  To minimise the number of false detections, a lower limit for the length of blue-shifted structures is set at $\sim$725~km, 17 pixels, 
and any thin blue-shifted feature with a length below this threshold is not counted.  
Like in Paper~I, we find that this length threshold limits the probability of false detections significantly.
A shorter length threshold allows small linear features in intergranular lanes to enter the sample which upon further scrutiny do not display the typical spectral character of RBEs. 
The output of the automated algorithm contains every detected event in every timeframe, and it also attempts to connect these single events over several timeframes by making chains of events which represents the time evolution of an RBE over several time steps.  If an event from one timeframe has 30\% of its pixels within a 3 pixel distance from any pixel in an event from the next timeframe,
these two events are considered to be the same RBE observed at different stages in its time evolution.  Because the lengths of events may vary greatly from frame to frame, the check has to be done in both temporal directions as the following event, which is supposed to link up with the current event, might be shorter than 30\% of the current event's length.  In this case, even if the following event lies completely on top of the current event, it would not be recognised as a valid link, while when performing the check "backwards" there is a 100\% overlap and the link is approved.  This check is performed on all events in a timeframe using all the events from the next timeframe and also the frame after that, meaning it allows for a jump in the chain of one timestep thus accounting for occasional moments of bad seeing.  
This automated method of connecting RBEs in time is an improvement compared to Paper~I, where the lifetimes of RBEs were estimated manually by looking at a subset of 35 separate events and following their existence in following time steps.  

Having the chains made automatically gives a large sample from which one can study the time evolution and make accurate statistics of the typical lifetimes of RBEs.  The cases where RBEs are connected in chains are also less likely to be false detections, as it would take errors to occur very close together spatially and temporally.  When it comes to the selection threshold for determining the temporal connection in chains, the threshold values were chosen based on trial and error.  Allowing for events to be further away from one another introduced a significant number of wrong connections, particularly in the regions where the RBE detections are densest, and by constraining the percentage of pixels that has to be within the selected distance, too many good connections were disallowed.  We found that the chosen parameters resulted in the most reliable statistics.

Once the chains are found, these can be used to get a measure of the transverse displacement and velocity.  This is achieved by finding the average orientation of all events in a chain and determine its perpendicular direction.  The transverse distance is then measured by taking the middle point of each event and projecting it onto the normal which allows for both a measurement of the distance traversed transversally from one event to the next and also the velocity at which the RBE is moving transversally as we know the time difference between the two events.
As explained above, there is an upper limit of 3 pixels on the spatial separation between RBEs in subsequent time frames. Combined with the cadence of 11.8~s, one would expect to have an upper limit for the transverse velocity of 11~\kms. However, this is not the case since the transversal displacement is measured from the RBE midpoints along the normal to the average direction of the RBE and this can give a larger separation than 3 pixels. 

We have made extensive use of CRISPEX 
\citep{vissers2012crispex} 
a widget based analysis tool programmed in the Interactive Data Language (IDL) that allows for efficient exploration of multi-dimensional datasets. 
With CRISPEX, the \Halpha\ and \CaIIIR\ datasets can be examined at the same time and the tool was particularly useful for the verification of the RBE identifications from the automated detection method. 

Figure~\ref{fig:rbedetail} shows details of one RBE event, clearly identified in both \Halpha\ and \CaIIIR. Panels (e) and (f) show the RBE as dark streaks against a background of granulation in \Halpha\ and inverse granulation in \CaIIIR. For the spatial locations marked with asterisks, the spectral time evolution is shown in the top panels. This particular RBE displayed little sideways motion so that a single pixel location shows the actual spectral evolution of the RBE. These panels illustrate the defining characteristic of RBEs: a short-lived asymmetry in the blue wings of these strong absorption lines. The bottom panels show the maps used in the automated detection process: in \Halpha\ the disturbing background is attenuated by subtraction of the corresponding far blue wing image. Through this subtraction, the granulation intensity pattern is inverted so that intergranular lanes, narrow linear features, become bright.  Our detection method is based on the negative of this difference images and searches for bright linear structures, removing the risk of intergranular lanes becoming false detection as these are now dark. 
In \CaIIIR, a Dopplermap is used and RBEs are identified as bright linear structures against a low-contrast background. The detected structures are marked as thin black lines.
Panels (c) and (d) show the detailed spectra at the time and location of the asterisks in panels (e) and (f) as solid lines. To separate the RBE spectral profile from the background profile, a reference spectrum constructed from averaging over the whole FOV is subtracted, resulting in the dashed profile. 
This RBE profile is clearly a blue-shifted component separated from the background spectrum, a characteristic that is also apparent from its dynamical evolution which appears to be unrelated to the chromospheric background seen closer to line centre.
The RBE profile is used to determine the Doppler velocity and Doppler width, following the method of Paper~I.
For this location in the RBE, the \Halpha\ Doppler velocity is 39~\kms\ and width 16~\kms, 
the \CaIIIR\ Doppler velocity is 23~\kms\ and width 13~\kms. 

The length of an RBE was determined using the same method as Paper~I: by first extending the detection skeleton (the thin black line in panels (g) and (h) outlining the detected event) by 20 pixels in both directions and dilation in width and then determining the points where either the Doppler velocity or Doppler width of the RBE profile is 0. 
This definition is equivalent to measuring the length for which a feature shows enhanced absorption compared to the average spectral profile. 
Hence, detected events have a possibility of becoming shorter than the minimum length threshold set by the detection routine since the measured Doppler width can become 0 even though there is signal in the detection map. 

\section{Results}

\subsection{Number of RBE detections}

\begin{figure}[!t]
\includegraphics[width=\columnwidth]{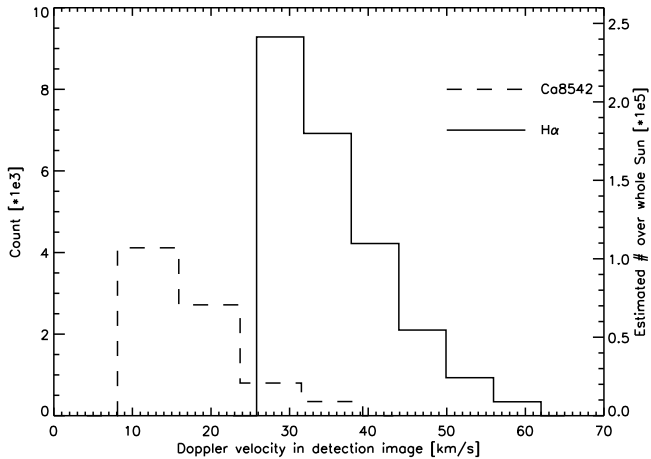}
\caption{Histograms of the number of detected events as a function of the Doppler velocity used in the detection image. \Halpha\ detection numbers in solid lines, \CaIIIR\ in dashed lines. The right axis scale indicates the estimated total number of events at any time over the whole surface of the Sun.}
\label{fig:nRBEvsVel}
\end{figure}

We found that the number of RBE detections varies significantly with the Doppler velocity from which the detection maps are constructed. 
Figure~\ref{fig:nRBEvsVel} shows histograms for the number of detected events as a function of the Doppler velocity in the detection  map. 
For both spectral lines, we see a significant increase in the number of detected events for lower Doppler velocities.
This trend is most pronounced for \Halpha, where we measure an increase from 341 at 63~km~s$^{-1}$ to 9285 events at 30~km~s$^{-1}$.  
We manually verified a large subsample of the events detected at 30~\kms\ and we confirmed that almost all events are genuine RBEs. 
For \CaIIIR\ we see a raise in the number of detected events from 346 at 35~km~s$^{-1}$ to 4116 at 12~km~s$^{-1}$. 
Manual inspection at 12~\kms\ however revealed a high fraction of false identifications. 
We conclude that this line position cannot be trusted for the automated detection of RBEs.
For the remainder of this paper, we present results from \CaIIIR\ RBEs detected at 20~km~s$^{-1}$, where we find the highest number of detections (2717 events). 
For \Halpha, we present results from RBEs detected at 45~km~s$^{-1}$ (2099 events) which gives a sample of similar size as for \CaIIIR. This choice is rather arbitrary since we regard all Doppler velocities reliable for RBE detection. We found it practical to have similar sample sizes for the two lines for the process of linking up the \Halpha\ and \CaIIIR\ RBEs.

The right axis scaling of Fig.~\ref{fig:nRBEvsVel} gives an estimate of the total number of RBEs at any time on the Sun. This estimate is derived from the average number of detections per time step extrapolated to the whole surface of the Sun.

\subsection{Statistical properties of RBEs}

\begin{figure*}[!ht]
\includegraphics[width=\textwidth]{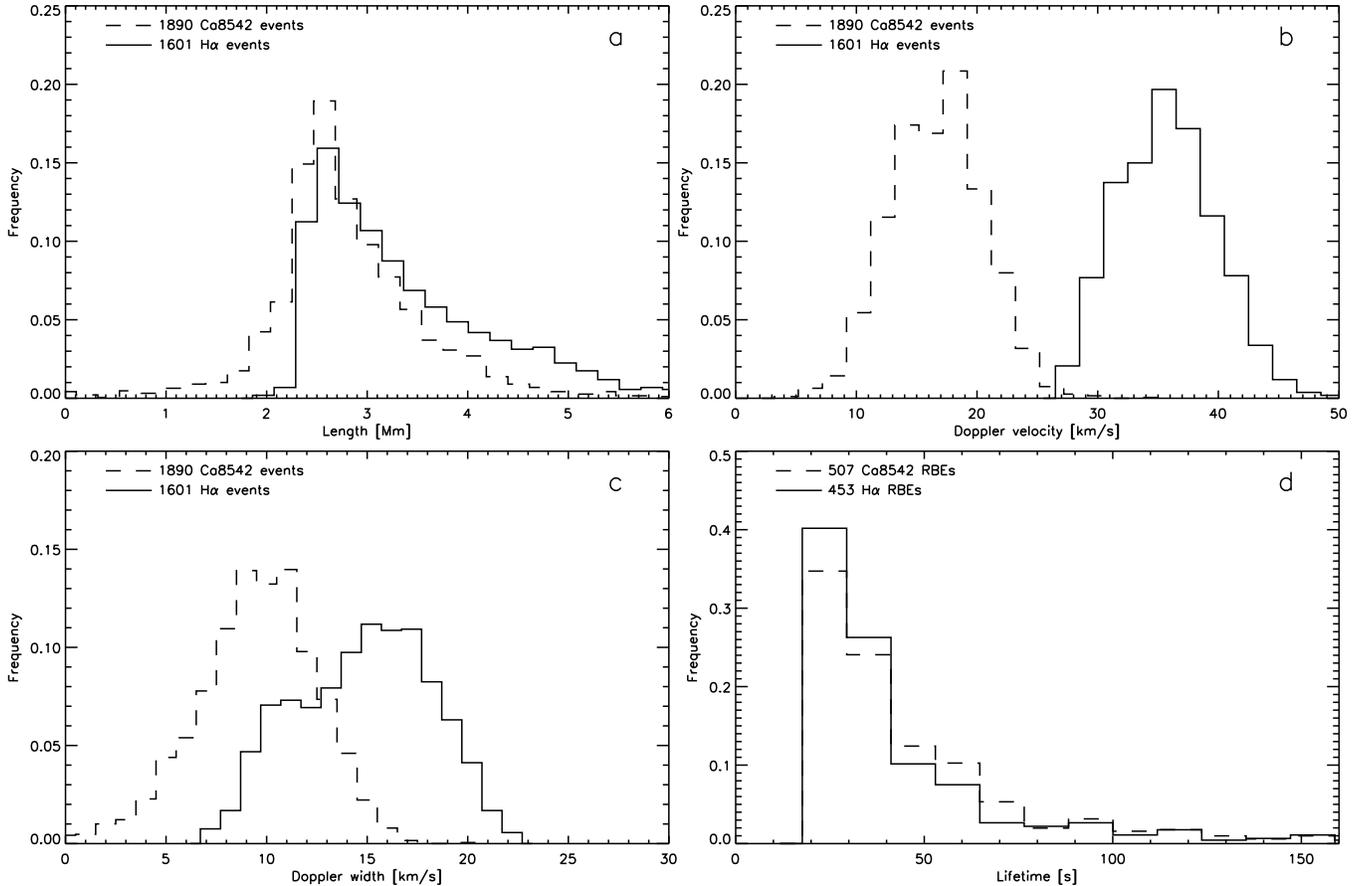}
\caption{Histograms for (a) length, (b) Doppler velocity, (c) Doppler width, and (d) lifetime of RBEs that were identified for more than 1 time step. Solid lines are \Halpha\ RBEs, dashed lines \CaIIIR\ RBEs. The histograms for Doppler velocity and width are average measures for each time step in each event. }
\label{fig:allchainshist1}
\end{figure*}

\begin{figure*}[!ht]
\includegraphics[width=\textwidth]{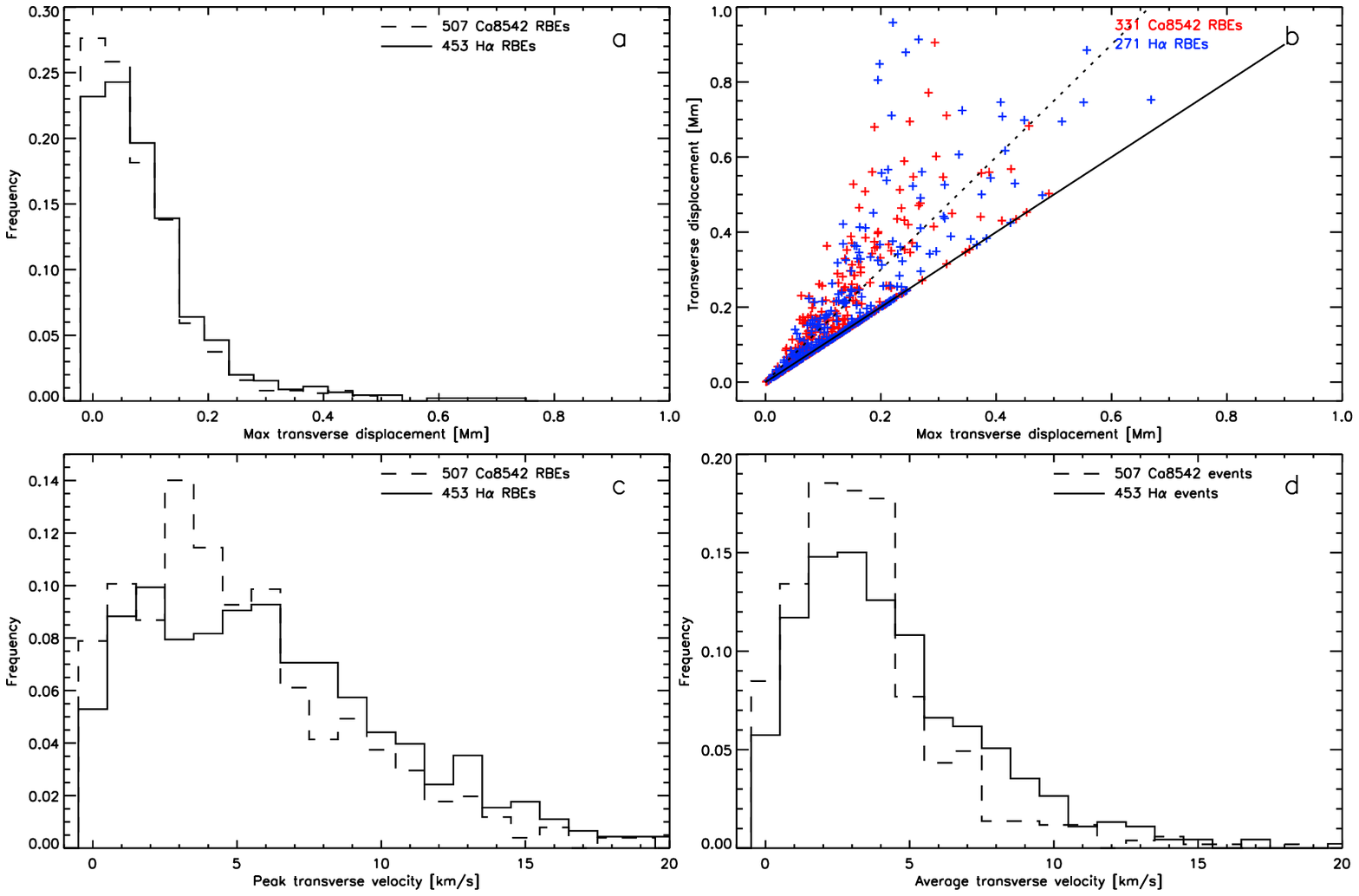}
\caption{Histograms for measurements of the transverse motion of RBEs: (a) maximum transverse displacement or distance between the two most extreme positions during the lifetime of an RBE, (c) maximum transverse velocity of all measured velocities during the RBE lifetime, and (d) all measured transverse velocities over all time steps. The scatter plot in panel (b) shows the total transverse distance covered by an RBE during its lifetime against the maximum transverse displacement of (a). The solid line marks the line where the two measures are equal, the dotted line marks where the total transverse displacement is 50\% larger than the maximum transverse displacement.}
\label{fig:allchainshist2}
\end{figure*}

Figure~\ref{fig:noevents}d shows a density map of the spatial locations of the 2099 \Halpha\ RBEs detected at 45~km~s$^{-1}$ drawn on a Stokes V magnetogram. 
Most RBEs are found around the patch of unipolar magnetic field in the upper right part of the FOV, the region that is dominated by fibril structures in the line core images and bright points in the wideband image.
There is another notable concentration of RBE detections below the centre which is associated with a small concentration of enhanced stokes V signal, and another smaller concentration to the left edge of the FOV. 
Besides these RBE concentrations there are a number of isolated RBEs scattered over the FOV that are not obviously associated with patches of stokes V signal. 
As explained in Sect.~2, the diagnostic potential of the Ca II 8542 -600 m\AA\ magnetograms is rather limited and it is well conceivable that a more sensitive polarization diagnostic would reveal association of the more isolated RBE signals with areas of enhanced polarization.  
 
Of the 2099 \Halpha\ events, 498 were detected only in one timeframe, while the rest linked up to other events in nearby timeframes to create a total of 453 RBEs existing over several sequential time steps. We will call the events that could not be linked up with events in other time steps ``single-frame RBEs'', and the others ``multi-frame RBEs''.

The map of \CaIIIR\ events (Fig.~\ref{fig:noevents}b) shows a similar distribution of most RBEs being associated with the magnetic field concentrations and a number of events scattered over the the FOV. 
Of the 2717 \CaIIIR\ events, 827 were single-frame events, and the rest was linked up as 507 temporally resolved or multi-frame RBEs.

Most of the multi-frame and single-frame RBEs were located in regions with enhanced polarization signal, and of the isolated RBEs that lie scattered over the FOV, the majority are single-frame RBEs, as can be derived from Fig.~\ref{fig:noevents}b and d.

For each individual detected event the Doppler velocities and Doppler widths were calculated at each pixel in an event.  
The position of the pixels and timeframe in which it is found were used for a measure of lengths and lifetimes of the RBEs.  
Figure \ref{fig:allchainshist1} and \ref{fig:allchainshist2} display histograms of all the important properties of the 1890 and 1601 events that were found to make up 507 and 453 multi-frame RBEs in \CaIIIR\ and \Halpha, respectively.  
The single-frame events are not included in this statistic as they lack time evolution, and thus cannot contribute to some of the histograms, such as transverse motions.

In panel (a) of Fig.~\ref{fig:allchainshist1} the lengths of all events are shown for both \CaIIIR\ and \Halpha. 
Typically, the \Halpha\ RBEs are found to have lengths 2$-$6~Mm, with an average of 3.5~Mm, while the \CaIIIR\ RBE lengths are seen to lie between 1.5 and 4.5~Mm with an average of 2.9~Mm.
A few \CaIIIR\ RBEs were found to have lengths shorter than the detection threshold of 725~km which is possible since the method of measuring the RBE length is different from the RBE detection (see Sect.~\ref{sect:method}).
No short \Halpha\ RBEs were found and we see a sharp lower cutoff at 2~Mm.

Panel (b) of Fig.~\ref{fig:allchainshist1} shows the average Doppler velocity of all RBEs. 
The average RBE Doppler velocities lie between 10--25~km s$^{-1}$ for \CaIIIR\ and 30$-$50~km s$^{-1}$ for \Halpha, with the mean value of the distribution of average Doppler velocities at 17.7 
 and 36.8~km s$^{-1}$, respectively. 
The maximum of all Doppler velocity measurements is as high as 50~km s$^{-1}$ for \CaIIIR\ and 62~km s$^{-1}$ for \Halpha.

Panel (c) of Fig.~\ref{fig:allchainshist1} shows the average Doppler width of all RBEs. 
As for the Doppler velocity, the Doppler widths are noticeably different in \CaIIIR\ and \Halpha\ with averages of 10~\kms and 15.3~\kms, respectively, and the range going from 0 to 17~\kms\ in \CaIIIR, while \Halpha\ RBEs display Doppler widths from 7 to 23~\kms.

Panel (d) of Fig.~\ref{fig:allchainshist1} shows the lifetimes of the RBEs. 
There is no major difference between the lifetime of \CaIIIR\ and \Halpha\ RBEs and they both have lifetimes varying 
from 24~s 
up to 2.5 min. 
There were a few detections of very long lived RBEs with lifetimes of more than 3 min. They were all found in regions with a high density of RBEs and visual inspection exposed these events as falsely connected chains of several separate multi-frame RBEs. 

Figure~\ref{fig:allchainshist2} show different measures of the transverse motion of RBEs. 
Panel (a) shows a histogram of the maximum transverse displacement which is the distance between the two most extreme transverse positions of the RBE during its lifetime. 
All RBEs have some transverse displacement and several of the RBEs display transverse displacements of more than 400~km.
The scatter plot in panel (b) shows the integrated transverse displacement, or the sum of the absolute values of each displacement during the RBE lifetime, against the maximum transverse displacement.
The dotted line marks where the integrated transverse displacement is 50\% larger than the maximum transverse displacement, the RBEs that lie above this line display considerable back-and-forth swaying motion during their lifetime. 
The RBEs that lie close to the solid line are mainly moving in one direction. 
Panel (c) shows the maximum transverse velocity for each RBE 
and panel (d) shows all measured transverse velocities. 
Transverse motions can be at velocities as high as 20~\kms\ 
although most events' velocities lie below 15~\kms\ with the bulk of the peak velocities around 5~\kms.

We observe many RBEs displaying back-and-forth sideways swaying motion during their life time. 
For all the RBEs that have a duration of more than two time steps, we measured how many times there is a change in direction. 
We found that for both \Halpha\ and \CaIIIR\ RBEs, the change in direction happens on average 1.7 times during their lifetime.

\begin{figure*}[!t]
\includegraphics[width=\textwidth]{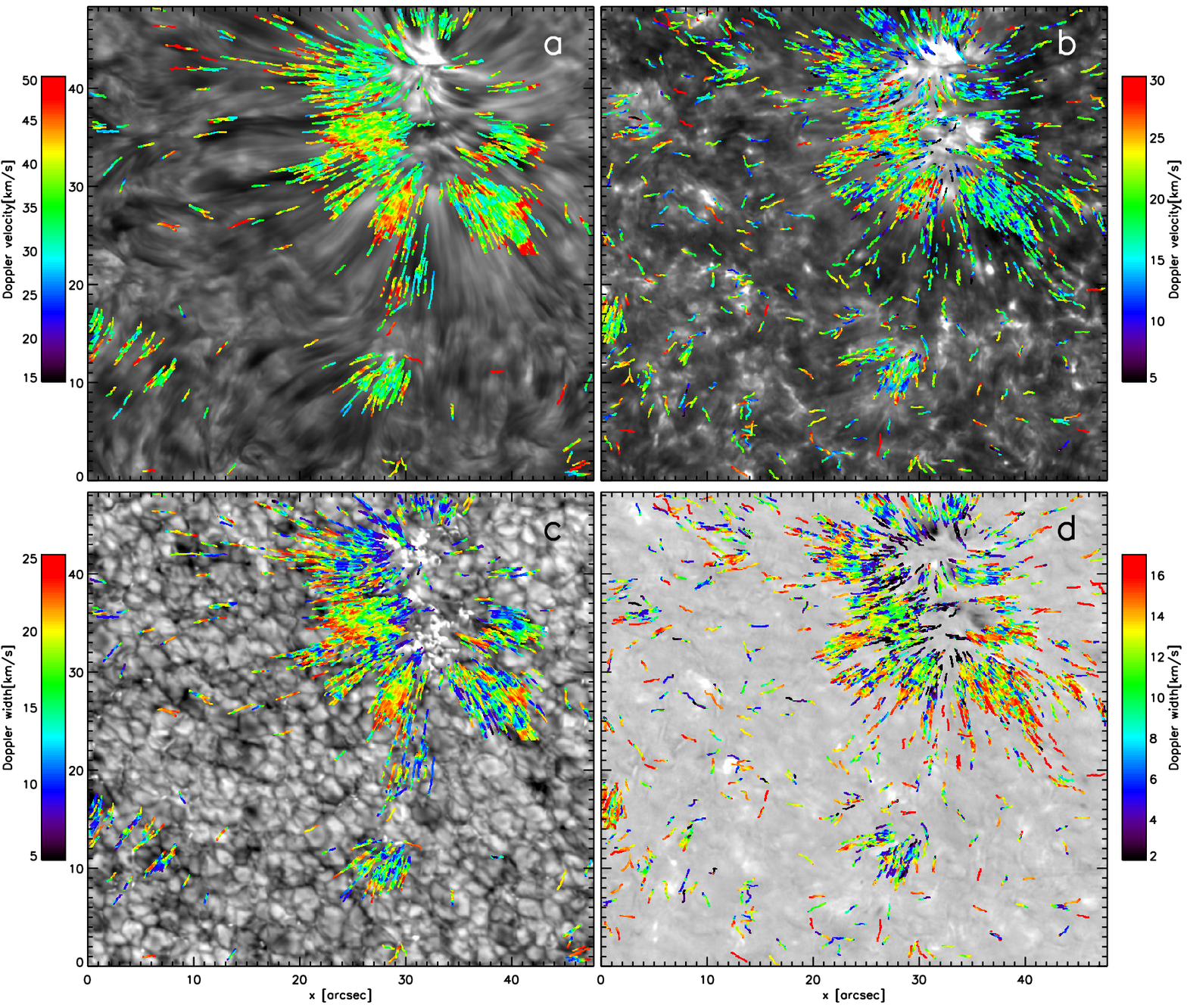}
\caption{Doppler properties of RBEs as a function of their position in the FOV, \Halpha\ RBEs at left, \CaIIIR\ RBEs at right. Color-coded Doppler velocities (top panels) and Doppler widths (bottom panels) drawn on background images of (a) \Halpha\ line centre, (b) \CaIIIR\ line centre, (c) \Halpha\ wideband, and (d) \CaIIIR\ $-20$~\kms\ detection map.}
\label{fig:vel_width}
\end{figure*}

\begin{figure}[!h]
\includegraphics[width=\columnwidth]{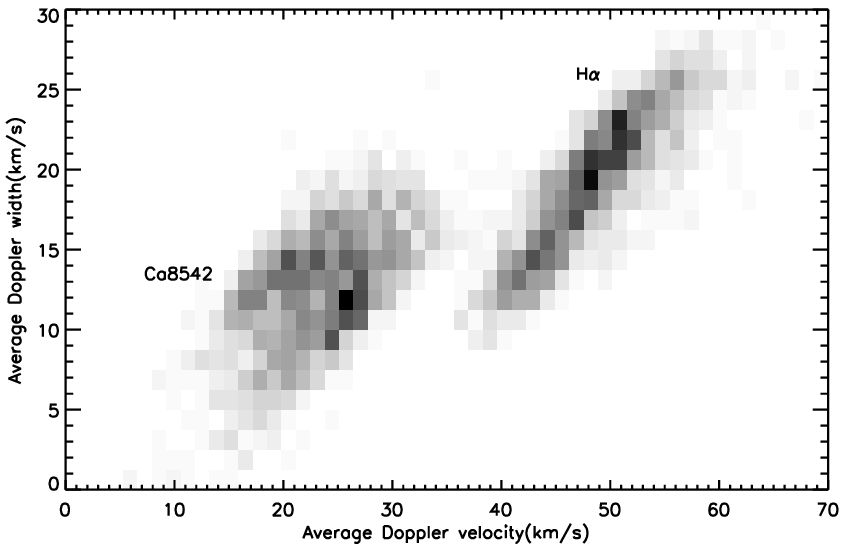}
\caption{Scatter plot of average Doppler width vs. average Doppler velocity of all RBEs, represented as a 2-dimensional density functions. Darker regions represent the densest areas with data points.}
\label{fig:scatter_width_vs_vel}
\end{figure}

Figure~\ref{fig:vel_width} shows the variation of the Doppler velocity and Doppler width for RBEs along their length as color-coded lines drawn at their spatial location in the FOV. 
All detected RBEs are drawn in successive order, with RBEs detected early in the time series drawn first so that they are covered by later RBEs that occur at the same spatial location. 
The RBE occurrence in the target area is dominated by the magnetic region which gives a clear sense of the orientation of the RBEs and where they originate. 
This representation gives a clear visual impression of many RBEs increasing in both Doppler velocity and Doppler width towards their top. 
This trend is most clear for the \Halpha\ RBEs. 
It should be noted that despite the visual trend, there is a significant number of RBEs that have a more erratic variation of Doppler properties along their axis and even examples of decreasing trends. 

There is a clear correlation between increased Doppler width for increasing Doppler velocity, this is illustrated in 
Fig.~\ref{fig:scatter_width_vs_vel}.

\subsection{The RBE \CaIIIR-\Halpha\ Connection}
\label{sec:CaHaConnection}

\begin{figure*}[!t]
\includegraphics[width=\textwidth]{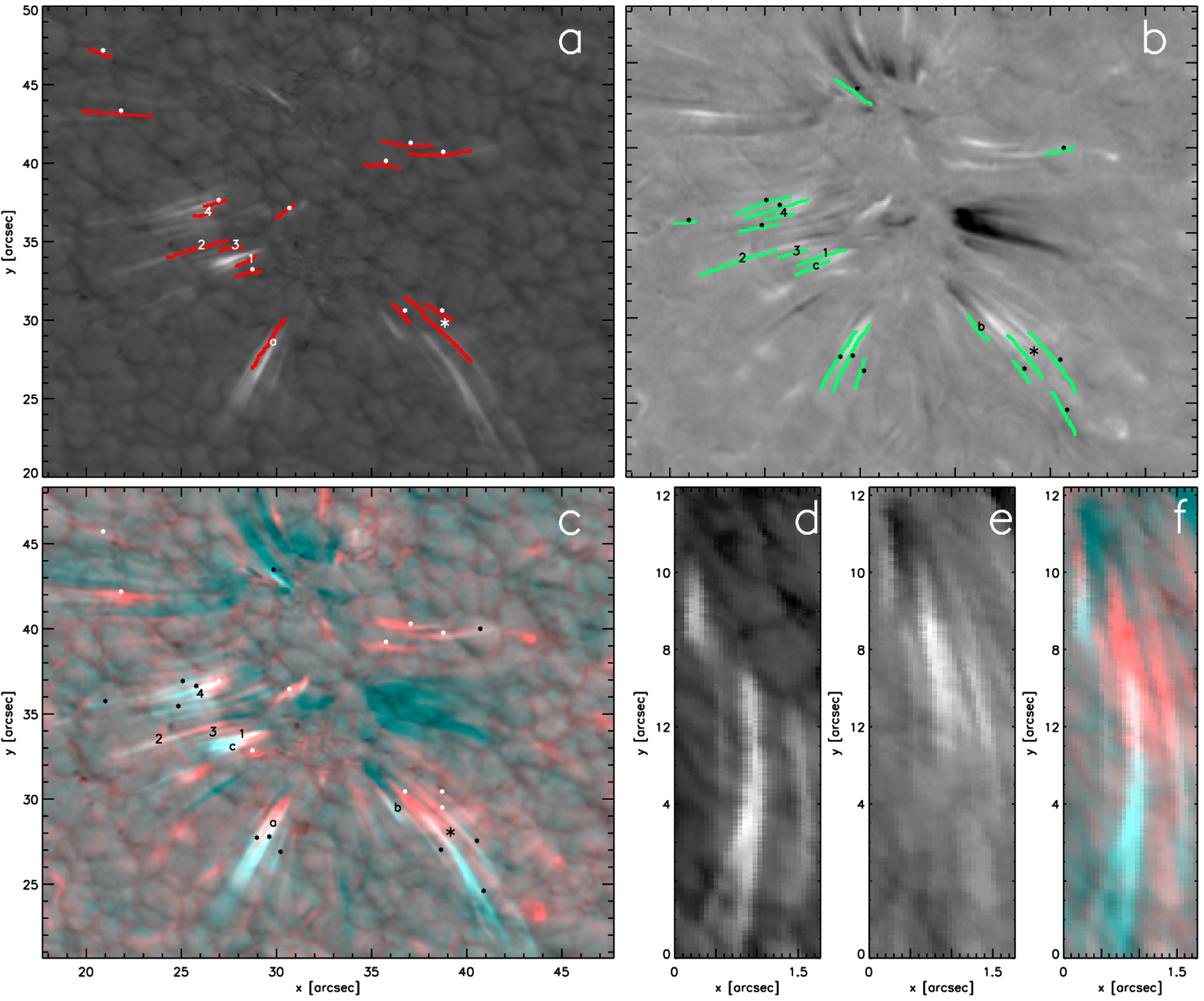}
\caption{Examples of RBE detections. (a): \Halpha\ $-45$~\kms\ map used for RBE detection, RBEs are white. Red lines mark \CaIIIR\ detections from the same time step. (b): \CaIIIR\ $-20$~\kms\ detection map, RBEs are white. Green lines mark \Halpha\ detections from the same time step. (c): composite image of the \Halpha\ and \CaIIIR\ detection maps: the \CaIIIR\ image filled the red channel, the \Halpha\ image filled the blue and green channels of the true color RGB image. Detections marked 1--4 are RBEs connected in both \Halpha\ and \CaIIIR\ for this time step, detections marked a--c are RBEs connected to the other spectral line in a different time step, detections marked with stars are RBEs that are not identified in the other line. The three bottom right panels show a closer view of one RBE marked with a larger star at $(x,y)=(39,28)$: (d) the \Halpha\ detection map, (e) the \CaIIIR\ detection map, and (f) the \Halpha--\CaIIIR\ composite image.}
\label{fig:hacaswitched}
\end{figure*}

From the initial detection of 2717 and 2099 events in \CaIIIR\ and \Halpha, respectively, a total of 1890 and 1601 events combined to form 507 and 453 multi-frame RBEs.
These chains from both spectral lines were then compared to one another in order to see if there is a connection between the \CaIIIR\ and \Halpha\ RBEs.
The same method that was used for linking separate events to multi-frame RBE chains was employed to link the chains in the two spectral lines. 
The automated method managed to find about 1/3 of the multi-frame RBEs in both spectral lines simultaneously resulting in 169 \CaIIIR-\Halpha\ RBE chains.
Figure~\ref{fig:noevents}c shows the distribution of these 169 chains, drawn on a wideband image. The \Halpha\ parts of the chains are drawn as green lines, the \CaIIIR\ parts as red lines, and the region where they overlap is drawn in yellow.

Our detection technique finds that a large fraction of the RBEs are detected and matched up in both lines, with a small fraction not matching up. However, from visual inspection of the dataset, it becomes clear that for virtually all RBEs detected in one spectral line a corresponding RBE signature is observed in the other spectral line.  This is illustrated in Fig. \ref{fig:hacaswitched}a and b, where close-ups of one time step are shown for both lines with the detected events in the other spectral line over-plotted.
The automated detection method managed to link 7 RBEs in both spectral lines: 4 are found in both lines for this time step (marked as 1,2,3, and 4), 
1 \CaIIIR\ RBE is linked to an \Halpha\ RBE in another time step (marked 'a'),  
and 2 \Halpha\ RBEs are linked to \CaIIIR\ RBEs in another time step (marked 'b' and 'c').
The other detected RBEs were not flagged as connections (marked with small stars).
Visual inspection however suggests a much higher connection rate which becomes even more apparent in the composite map in panel c. 
Here, \Halpha\ and \CaIIIR\ detection maps are combined in a color image where bright \Halpha\ features are blue-green, and bright \CaIIIR\ Doppler features are red. Regions where these features overlap (i.e., overlapping RBEs) become white. 
This map suggests that all RBEs have signal in both \Halpha\ and \CaIIIR. 

Panels (d), (e), and (f) of Fig.~\ref{fig:hacaswitched} offer a close-up centred on one of the longer RBEs. This long RBE was detected in both \Halpha\ and \CaIIIR (see panels (a) and (b)), but the automated method did not connect them. The images (and in particular the composite image) show that there is considerable overlap between the two lines. 
The other long but thinner RBE to the right is also detected in both lines but not flagged as connections -- the composite image leaves no doubt that the two detections are of the same event. 
And the small RBE in the upper left was only detected for \Halpha\ for this time step (marked 'b', connected to a \CaIIIR\ detection in the next time step) but there is indeed RBE signal present also in \CaIIIR.

Figure~\ref{fig:hacaswitched} demonstrates that \CaIIIR\ RBEs have their lower part closer to the magnetic regions, while \Halpha\ RBEs extend to larger distances away from these regions. 
This trend of \CaIIIR\ RBEs being closer to the magnetic regions is also illustrated in Fig.~\ref{fig:noevents}c:  
most of the red lines (\CaIIIR) are at the base of the green \Halpha\ RBEs.
The single RBE in Fig.~\ref{fig:rbedetail} is also found in \CaIIIR\, with the latter feature located closer to the magnetic bright points in the upper right (where the RBE started to appear a few time steps earlier) than in \Halpha.

When comparing the statistics of RBEs connected in both spectral lines to the ones that are not connected, two of the properties stand out.  The average lifetime of an RBE seen in both spectral lines is 83.9~s
whereas in the RBEs only detected in one of the lines, the average lifetime is 45.6~s and 46.1~s for \CaIIIR\ and \Halpha, respectively.  
Also the average length of RBEs seen in both spectral lines, 3.1~Mm and 3.7~Mm for \CaIIIR\ and \Halpha, respectively, is larger than that of the RBEs only observed in one spectral line, 2.8~Mm and 3.4~Mm for \CaIIIR\ and \Halpha, respectively.
This suggests that a successful connection between the two lines is more likely for RBEs that are longer lived and have longer spatial extent. 

The statistics for Doppler velocities and Doppler widths for the connected RBEs shows the same trend as for RBEs detected in only one line: \Halpha\ RBEs have generally larger Doppler velocity and larger width.

\subsection{Scattered Events}
\label{sec:scattered_events}

\begin{figure}[!t]
\includegraphics[width=\columnwidth]{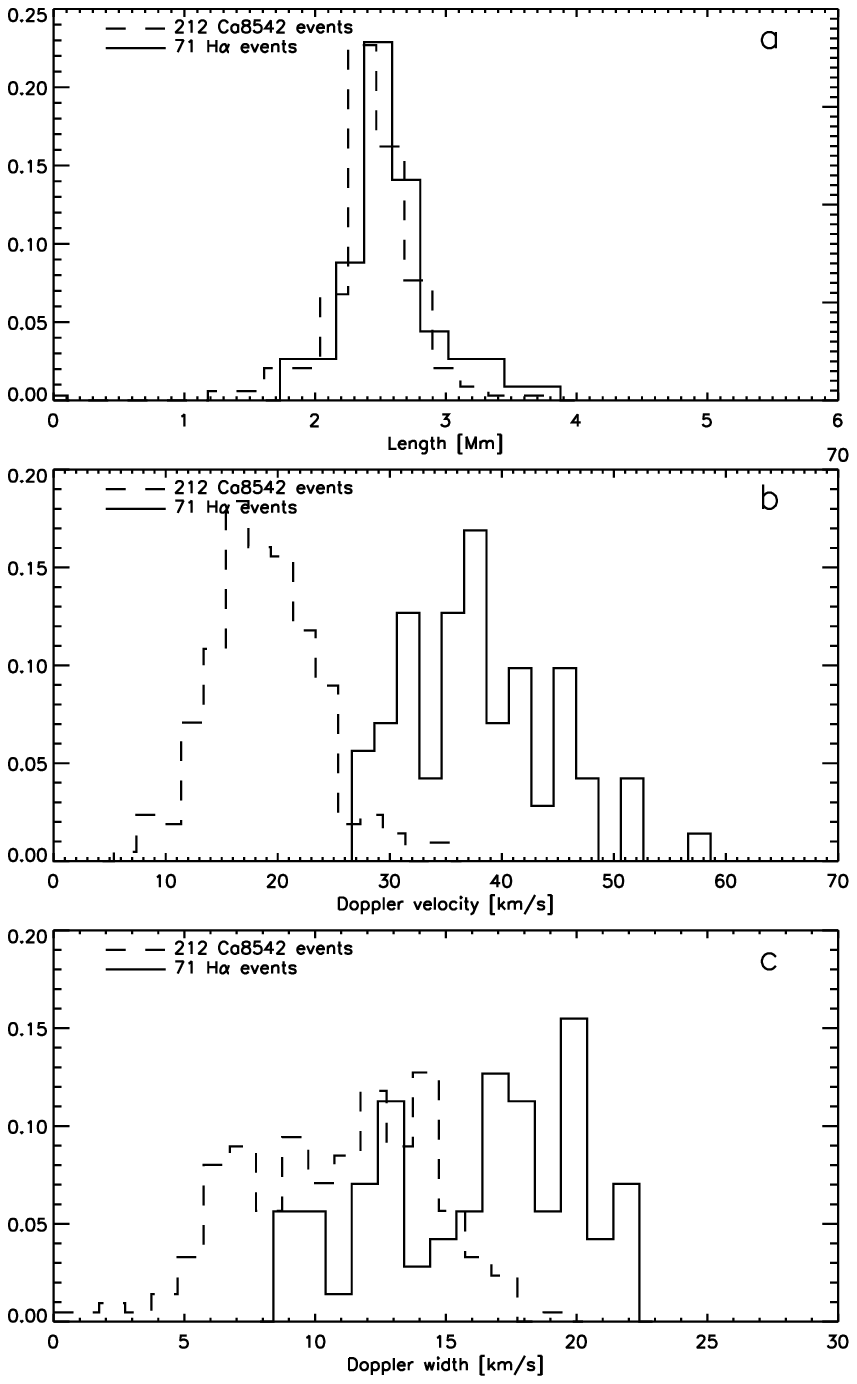}
\caption{Histograms for (a) length, (b) Doppler velocity, and (c) Doppler width for RBEs scattered over the FOV, i.e., not associated with enhanced polarization signal (see Fig.\ref{fig:noevents}d).}
\label{fig:scattered_hist}
\end{figure}

\begin{figure*}[!t]
\includegraphics[width=\textwidth]{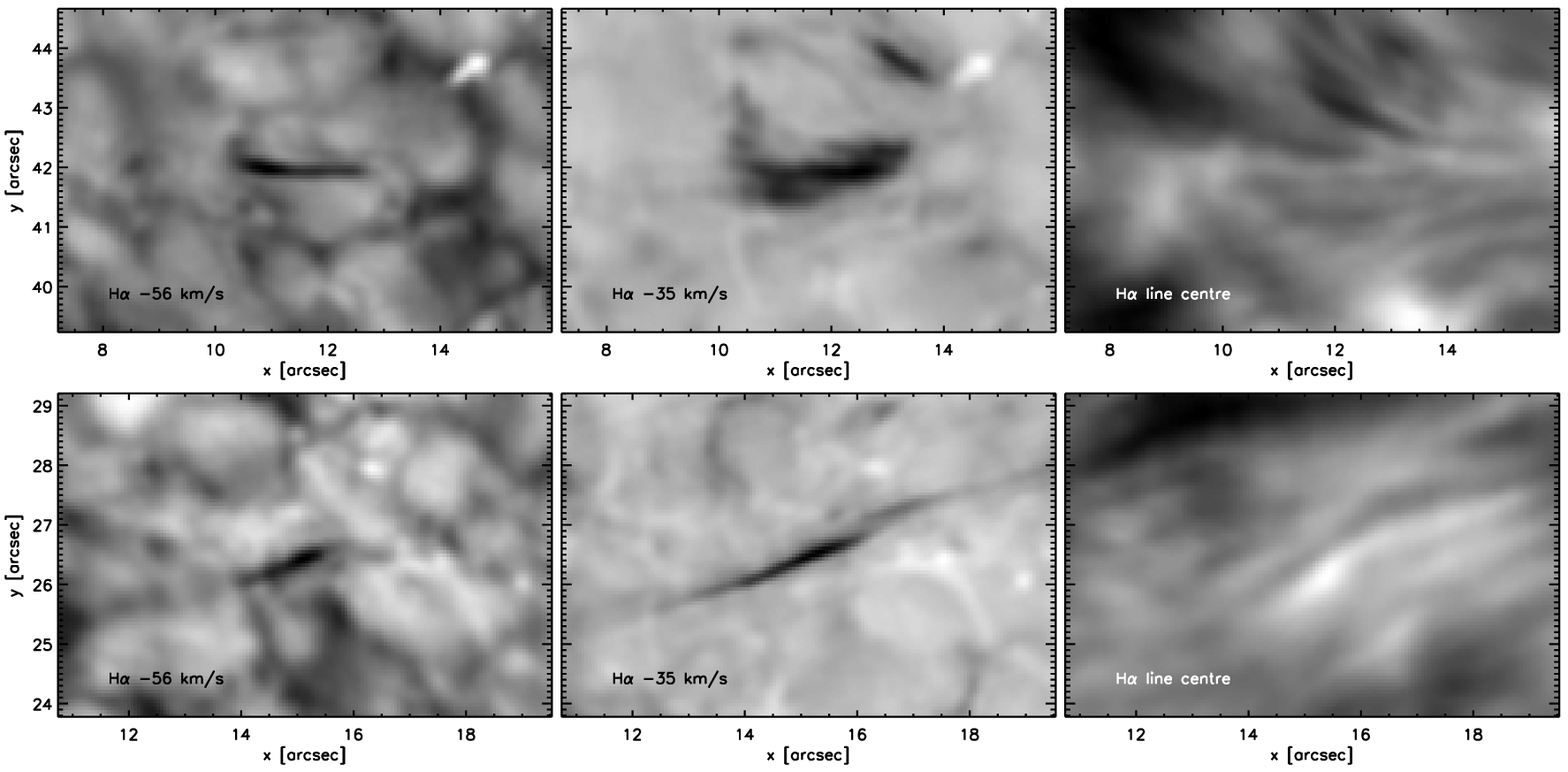}
\caption{Sample \Halpha\ images of two scattered RBEs. Left panels are images in the blue wing at $-1200$~m\AA, middle panels at $-760$~m\AA, and right panels in line centre.}
\label{fig:scattered_examples}
\end{figure*}

Figures~\ref{fig:noevents} and \ref{fig:vel_width} show that apart from the dense concentrations of RBEs associated with regions of enhanced polarization, there are a significant number of events scattered over the FOV.
Here we present statistical properties of these isolated events by masking out the three regions with densest RBE concentrations. 

Figure~\ref{fig:scattered_hist} shows histograms for length, Doppler velocity and Doppler width.
The histogram for RBE length peaks just below 3~Mm, just like for the multi-frame RBEs in Fig.~\ref{fig:allchainshist1}a, but there is no extended tail towards longer lengths
(note that some of the scattered RBEs show temporal evolution and are part of both Fig.~\ref{fig:allchainshist1} and \ref{fig:scattered_hist}).
The histograms for Doppler velocity and width are similar as for multi-frame RBEs.

Figure~\ref{fig:scattered_examples} shows sample images of two scattered \Halpha\ RBEs. 
The top panels are from an event that appeared one time step earlier in the far blue wing in the middle of a granule.
The RBE could be followed for 4 time steps so it has a life time of more than 36~s. 
In the left panel, the RBE appears to cross a granule, while in the images closer to line core (for example the middle panel), it seems to be part of a structure that is rooted in  the intergranular lane to the right of the granule. 
No photospheric bright point or enhanced polarization signal that is obviously associated with this event is present in this area.

The bottom panels show an RBE that was visible for 8 time steps, and like the previous example seems to cross granules and is not obviously connected to magnetic regions in the vicinity. 
The line core image however suggests that the RBE is associated with the dense fibrils that are rooted in the extended magnetic region to the upper right that dominated the FOV in Fig~\ref{fig:noevents}.
The closest photospheric bright points along the direction of the RBE main axis are more than 10\arcsec\ away.

\section{Discussion}
\label{sect:discussion}

We studied the disk-counterparts of type II spicules that are observed as short-lived, elongated features in the blue wings of strong absorption lines like the \Halpha\ and \CaIIIR\ spectral lines. 
These features are also known as rapid blue-shifted excursions due to their asymmetric, blue-ward spectral profile. 
The work presented here is in many ways a continuation and elaboration of the work of Paper~I. 
We observed a similar target, a coronal hole, and employed a program of simultaneous \Halpha\ and \CaIIIR\ spectral scans. 
The observation program was specifically tailored for the observation of RBEs with reasonably high temporal cadence and sufficient resolution for robust spectral analysis. 
This allowed us to more directly compare the two spectral diagnostics, a comparison which could only be done indirectly from the data employed in Paper~I, where RBEs were studied in two separate but sequentially recorded single-line time series in \Halpha\ and \CaIIIR. 
Furthermore, we have extended the method of automated detection of RBEs to allow for a temporal connection between RBE detections so that properties such as lifetimes and transversal motion could be studied more rigorously on an extended statistical sample. 
In Paper~I, these properties were measured manually on a limited subsample of the RBE detections. 

\subsection{Statistical properties of RBEs.}
We present histograms on RBE lengths, Doppler velocities and Doppler widths for 1890 \CaIIIR\ and 1601 \Halpha\ RBEs in Fig.~\ref{fig:allchainshist1}. 
These measurements agree very well with those presented in Paper~I.
We confirm that a larger number of longer RBEs ($\gtrsim$3.5~Mm) can be observed in \Halpha, while the shortest RBEs ($\lesssim$2~Mm) are observed in \CaIIIR.
We confirm that in \Halpha, one observes larger Doppler velocities and larger Doppler widths than for \CaIIIR\ RBEs.

These RBEs were connected in temporal chains and combined to form 507 \CaIIIR\ and 453 \Halpha\ multi-frame RBEs for which we could determine the life time and transversal motion. 
This is an increase in statistical sample size of more than an order of magnitude for this type of measurements. 

Pereira et al. (private communication) 
measure statistical properties of 111 type II spicules from two {\it Hinode} Ca~II~H limb time series.
Their results allow for a more detailed comparison between RBE properties and spicule type II properties than was possible for Paper~I.
They find life times ranging between 20--160~s, with the peak of the distribution around 1~min and an average of about 80~s. 
For our sample of RBEs, we find a similar range in life times but the peak of the distribution well below 1~min and an average of 50~s.
We note that for the RBEs that can be identified in both \CaIIIR\ and \Halpha, we find an average life time that agrees well with the 
Pereira et al. 
result.

Pereira et al. 
find for the transverse motion the average velocity ranging between 0 and 20~\kms, with the peak of the distribution at about 6~\kms. 
We find a similar range in velocities, but most of the measurements for RBEs yield lower velocities, with the average at $\sim$4~\kms\ for \CaIIIR\ and $\sim$5~\kms\ for \Halpha. 
The measurements of the transverse motion of 35 RBEs in Paper~I, resulted in a transverse velocity of order 8~\kms. 
These \Halpha\ RBEs were selected manually but we confirmed that result by applying the method of automated measurement of transverse motion to the 2008 data set and found an average of 7.6~\kms\ for 179~\Halpha\ RBEs. 
We attribute this higher value for the transverse velocity to the higher cadence of the 2008 \Halpha\ data set which was almost two times faster than the 2010 data. 
This allows for the measurement of higher velocity motions because the upper limit of maximum allowed distance between time steps is larger.
For the 2008 \CaIIIR\ data set, which is at similar cadence as the 2010 data, we find an average of 4.4~\kms\ for 94 RBEs. 
For the 03-July-2010 data set, we find average transverse velocities that are very similar as for 27-Jun-2010. 

\citet{2007Sci...318.1574D} 
discovered Alfv{\'e}nic waves in the chromosphere using type II spicules as tracers for oscillatory motions. The velocity amplitude was measured for 94 spicules which can be compared to the distribution of the peak transverse velocity in 
Fig.~\ref{fig:allchainshist2}c. 
They
found a range between 0 and 30~\kms, with the peak of the distribution around 15~\kms. 
Like for the average transverse velocity, RBEs display a comparable range but mostly lower values for the peak transverse velocity as compared to type II spicules at the limb. 
Continuing on the arguments of the previous paragraph, we note that the measurements of the RBE transverse velocities seem to be limited by the cadence of the time series. 
Faster cadence observations are needed to properly resolve the distribution of transverse velocities of RBEs. 

For the transverse displacement, Pereira et al. 
find values ranging between 0 and 1.4~Mm, with most values between 0.1 and 0.5~Mm, in close agreement with the 
\citet{2007Sci...318.1574D} 
measurements. 
We find a total transverse displacement of less then 1~Mm for RBEs, with most RBEs covering a distance of less then 0.2~Mm during their life time. 

Even though the shapes of the statistical distributions of properties of RBEs and type II spicules observed at the limb differ in detail, the similarities between the distributions provide sufficient basis for the interpretation of RBEs being the disk counterparts of type II spicules, as already put forward in Paper~I.  
In addition to similarity in statistics, we observe that RBEs rapidly disappear or fade, and that RBEs show both swaying motion and apparent motion away from magnetic regions that can be interpreted as an upward motion. There are no signs of flows returning to the surface associated with RBEs. 
This kind of dynamical behavior is one of the defining characteristics of type II spicules observed in Ca II H at the limb. 

The most striking difference between RBEs and type II spicules is the distribution of lengths. 
Pereira et al. 
measure a range between 1.5 and 11.5~Mm, with the average length around 6~Mm. 
We find for RBEs an average length around 3~Mm and only few \Halpha\ RBEs measure up to 6~Mm.
This difference can be attributed to the fact that above the limb, faint signals towards the top of spicules can still be reliably detected against the dark background. 
Against the disk however, these top parts of spicules are likely to have too little opacity to be detected as an integral part of RBEs. 
So there is likely an intrinsic bias for shorter lengths measured for RBEs. 
For length measurements of type II spicules at the limb, there is arguably also a selection effect for detecting longer spicules that stick out above the dense forest of spicules encountered in limb observations. 
The same projection effect could possibly explain the trend of higher transverse velocity and transverse displacement measured for spicules at the limb: tall spicules that extend to greater heights move in a lower density environment, so for equal wave energy flux they can be expected to display higher velocities and thus cover greater transverse distances.

\subsection{RBE detection rate.}
In Paper~I, in the 40~min \CaIIIR\ time sequence with cadence 11~s, 413 RBEs were detected from Dopplermaps at 30~\kms. With the 215 time steps, this corresponds to a detection rate of 1.9 RBEs per time step. 
In the 24~min \Halpha\ time sequence with cadence 6.7~s, 608 RBEs were detected from difference maps at 60~\kms. With the 245 time steps, this corresponds to a detection rate of 2.5 \Halpha\ RBEs per time step. 
We note that detection velocities of 30~\kms\ for \CaIIIR\ and 60~\kms\ for \Halpha\ are on the extreme end of the average Doppler velocities found for RBEs, see Fig.~\ref{fig:allchainshist1} (or Fig. 10 in Paper~I). Given the typical values for the Doppler width and velocity it is likely that at these extreme detection velocities, a significant fraction of RBEs have absorption signals that are too faint to be detected by the automated method. 
In this work we investigated the detection rate as a function of the detection velocity used in the automated method. 
We find similar detection rates as Paper~I at the detection velocities used in Paper~I: 
3.4 RBEs per time step for \CaIIIR\ 
and 1.8 RBEs per time step for \Halpha.
Running the automated detection method on maps constructed at lower Doppler velocity, we find a significant increase in the detection rate, see Fig.~\ref{fig:nRBEvsVel}. 
At 30~\kms\ we had a detection rate of 48 \Halpha\ RBEs per time step, an increase in the number of detected events of more than a factor 25. 
For \CaIIIR\ we see a similar increase in the detection rate towards lower Doppler velocity but we found that at the lowest Doppler velocity possible in our data set, 12~\kms, the fraction of false detections was too large and the detections are not reliable. 
For the detection velocity for which we present RBE statistics, 20 and 45~\kms\ for \CaIIIR\ and \Halpha, we see an increase in the detection rate with a factor of about 6, to 14.2 and 11.1 RBEs per time step, as compared to the Paper I detection velocity. 
This resulted in a significant increase of the RBE sample used for detailed analysis in this work.

We find similar trends of increased RBE detections at lower Doppler velocity for the 03-Jul-2010 data set and the 15-Jun-2008 data sets used in Paper~I.

We also note that the automated detection method does not identify all RBEs present in the data. We estimate that for the 45~\kms\ \Halpha\ detection maps, only about half of the events that are visually identified as genuine RBEs are registered by the automated detection. 
Running the detection routine on lower Doppler velocity maps results in a lower relative number of undetected RBEs (about a quarter remains undetected) but we must underline that we considerably underestimate the abundance of RBEs in our data.

\citet{2011ApJ...730L...4J} 
proposed that type II spicules correspond to warps in two-dimensional sheet structures as an alternative to the main-stream interpretation of spicules being tube-like structures. 
One of their arguments in favor of sheet-like structures is that RBEs (which are hard to interpret as "sheets") are not the disk counterparts of type II spicules because the occurrence rate of RBEs is at least an order of magnitude lower than type II spicules observed at the limb.
From the number of detected RBEs in Paper~I, they extrapolate to an estimate of 10$^{5}$ RBEs on the Sun at any given time. 
This is then orders of magnitudes lower than estimates for type II spicules, $2 \times 10^{7}$ 
\citep{2010ApJ...719..469J}, 
and \citet{1968SoPh....3..367B}'s 
estimate of 10$^{6}$ for what is now considered as classical spicules.

The detection rate of RBEs at lower Doppler velocity we find here agrees with the estimate of a total number of  $\sim10^5$ RBEs (see Fig.~\ref{fig:nRBEvsVel}).
However, as argued above, the RBE detection rate we find here is an absolute lower limit that significantly underestimates the actual number of spicules for several reasons.

First, the automated detection rate rejects a significant fraction of real RBEs to avoid false positives. 

Secondly, there are clear indications that many spicules lack opacity to show a disk counterpart signal. Detecting an RBE-like feature against the disk requires observations of excellent quality, and most importantly requires that the RBE has a certain degree of enhanced opacity. Our disk observations indicate that the opacity of RBEs varies strongly between the \CaIIIR\ and \Halpha\ lines, with the former showing significantly fewer and shorter features (concentrated towards the bottom of the \Halpha\ features). This is compatible with an ever decreasing opacity in chromospheric lines as one moves away from the footpoints. This is not surprising because limb spicules also show a very strong decrease in intensity (with scale height of order 2000~km) over their full length. It is thus highly likely that there is a significant number of faint RBEs that cannot be detected even in \Halpha. We note that RBEs are very difficult to observe: detecting RBEs was impossible before the advent of subarcsecond, high cadence, high signal-to-noise imaging spectroscopy with Fabry-Perot interferometers. It is quite possible that even higher quality observations would increase the detection rate significantly.  

Thirdly, RBE detection is subject to strong Òvelocity filteringÓ: due to the unknown angle between the velocity vector and line of sight vector, our method always misses features that are significantly inclined from the line of sight. This is especially the case because we cannot distinguish RBEs from other chromospheric features at velocities closer to the core. The increase of the number of RBE detections for lower Doppler velocity as illustrated in Fig. 3 indicates that there might be a significant number of RBEs at even lower Doppler velocity that are not detected by our method. Detection of type II spicules at the limb suffer much less from these latter two effects: it is relatively easy to detect faint signals against the off-limb dark background (with space-based instruments like Hinode that do not suffer from seeing effects introduced by the Earth's atmosphere) and velocity filtering is much less an issue for the relatively wide pass-band of the Hinode/SOT Ca H filter.

In the discussion of Paper I, it was argued that the detection rate of RBEs on the disk correspond to a detection of 1.9 RBEs per linear arcsec if they were observed at the limb. This was considered as being in reasonable agreement with an order of magnitude estimate of the number of type II spicules per linear arcsec from the 
\citet{2007PASJ...59S.655D} 
observations. In this study, we confirm the results from Paper I and find that the detection rate extrapolates to $\sim 1-2\times10^5$ RBEs over the whole Sun. 
We caution that all of these occurrence rate estimates, including those of  
\citet{1968SoPh....3..367B} 
and 
\citet{2010ApJ...719..469J} 
are based on incomplete datasets, limiting assumptions, observational limitations, and should thus be taken with a grain of salt. For example, it is not clear from 
\citet{2010ApJ...719..469J} 
how deterministic their occurrence rate of $10^7$ type II spicules really is. The visual comparison in their paper is unconvincing with their simulations showing many more spicules than observations do:  $\sim 10^7$ type II spicules would correspond to more than 100 spicules per linear arcsec at the limb or more than 5 spicules per Hinode/SOT pixel, in contradiction to observations. This implies that the $10^7$ type II spicules estimated by 
\citet{2010ApJ...719..469J}
, is likely a significant overestimation of the number of type II spicules at the limb.  In any case, their argumentation for that number is based on a series of assumptions and models whose effect on the estimated number of spicules has not been studied or estimated. This sheds doubt on the robustness of their determined detection rate.

We conclude that the rejection of \citet{2011ApJ...730L...4J} 
of RBEs being the disk counterparts of type II spicules on the basis of too low occurrence rate is unjustified. 
There may well be sheet-like features in the atmosphere, but challenging the "tube hypothesis" based on rejection of the RBEs as disk counterparts of spicules is not convincing for the reasons outlined above.

\subsection{Acceleration along RBEs.}
Paper~I reported a systematic variation of the Doppler velocities and widths: many RBEs were found to show an increase of Doppler shifts and widths from the footpoint to the top. 
We confirm this trend in the 2010 data which is visually evident 
from Fig.~\ref{fig:vel_width} which shows the Doppler measurements as a function of their position in the FOV. 
In this representation, the orientation of (most of) the RBEs can be comprehended with respect to their footpoints which appear to be rooted in the magnetic field concentrations. 
The trend of increasing Doppler velocity and width for many RBEs is particularly striking for the \Halpha\ RBEs.
We note that there is also a significant number of RBEs displaying a more erratic variation of their Doppler properties. 
%

Irrespective of the variation along the RBE main axis, we report a more general trend of increasing Doppler width for Doppler velocity in Fig.~\ref{fig:scatter_width_vs_vel}.
This trend illustrates the characteristic spectral profile at the location of RBEs which consists of a wide blue-ward asymmetry (from which the RBE profile is then extracted by subtraction of a background profile). 
This is different from profiles with a narrow, isolated absorption component which can for example sometimes be observed for on-disk coronal rain 
\citep{2012ApJ...745..152A, 2011RhodosAntolin}. 
Sometimes coronal rain condensations appear as separated absorption profiles in the far wings of \Halpha. 
This is unusual for RBEs, even for RBEs that are oriented along the line-of-sight and appear as small, roundish features with strong absorption in the blue wing (dubbed "black beads" in Paper~I), the spectral profile displays enhanced absorption throughout the blue extending to large Doppler offset. 

In any case, the correlation between Doppler velocity and widths and tendency towards increased values for both parameters towards the top of the RBEs may suggest a link between the acceleration and heating mechanism involved in RBEs, if we speculate that the Doppler widths are related to temperature. Whatever the cause of this intriguing correlation, it provides strict constraints for any theoretical model for spicules.

\subsection{Comparing \Halpha\ and \CaIIIR\ RBEs.} 
From the sequentially recorded \Halpha\ and \CaIIIR\ data sets of Paper~I, it was observed that \CaIIIR\ RBEs are located closer to magnetic regions and it was inferred that the \CaIIIR\ line samples the lower part of spicules. 
In our data sets with co-temporal \Halpha\ and \CaIIIR\ observations, we can directly confirm that the two lines sample different parts of RBEs, with the \CaIIIR\ line sampling the bottom part of the spicule, and \Halpha\ the upper part, and both lines sampling an overlapping region in the middle (see Fig~\ref{fig:hacaswitched}).
Even though the automated detection method does not nearly identify all RBEs in both lines, we find from visual inspection that virtually all RBEs display signal in both lines.  

This observation strengthens the explanation for the differences in Doppler velocity measured in \Halpha\ and \CaIIIR\ (see Fig.~\ref{fig:allchainshist1}) as put forward in Paper~I: \Halpha\ is sampling the top part of spicules where lower density plasma is propelled to higher velocities.

The observation of the \CaIIIR\ and \Halpha\ lines sampling different parts of RBEs also removes some of the discrepancy between length measurements of RBEs and type II spicules at the limb: we can infer that the actual length of the RBEs is longer than measured in the two lines separately.
In addition, of course, it remains the case that we cannot expect to completely remove the length discrepancy between RBEs and type II spicules. This is because off-limb measurements allow for more reliable measurements of faint signals.

\subsection{Scattered RBEs.}
We find most RBEs to be concentrated near regions with enhanced magnetic fields. 
In addition to these dense concentrations, we find a number of more isolated RBEs that are scattered around the FOV.
When comparing statistical properties of length, Doppler velocity and Doppler width, we find that these isolated events are not particularly different from the RBEs in denser regions, except that we do not find a population of longer RBEs. 
For these isolated RBEs, it is often unclear if they have any association with regions of enhanced magnetic fields, although we note that our diagnostic, \CaIIIR\ $-600$~m\AA\ Stokes~V maps, is not a particularly sensitive measure for magnetic fields. 
One might speculate that a more sensitive diagnostic might reveal a connection to weaker and possibly more extended, magnetic fields.

From fixed wavelength, $-1300$~m\AA\ \Halpha\ observations from the 1.6 m New Solar Telescope (NST) in 
Big Bear Solar Observatory, 
\citet{2010ApJ...714L..31G} 
report the observation of small jet-like features that appear to originate from intergranular lanes. 
The authors claim that these features are different from RBEs, as they are not unequivocally tied to strong 
magnetic field concentrations.

Figure~\ref{fig:scattered_examples},
shows blue-wing \Halpha\ images of two examples of scattered RBEs. 
The left panels show far blue wing images of these events. 
In these images, these events seem to match the description of the jet-like features described by 
\citet{2010ApJ...714L..31G}: 
narrow features that cross granules and are not obviously connected to magnetic regions in the vicinity. 
The spectro-temporal information that we have for this data however reveal that these events are RBEs. 
Even though the bottom example of Fig.~\ref{fig:scattered_examples} is not directly associated with photospheric bright points, it seems plausible that this feature is related to the extended magnetic region in the upper-right part of the FOV. 

\citet{2010ApJ...714L..31G} 
report a typical length of 1~Mm and a typical width of 0.2~Mm for the jet-like features in their blue-wing \Halpha\ data. 
The width agrees well what we find for RBEs, the length is compatible but on the short side. 
It should be noted that the length in 
\citet{2010ApJ...714L..31G} 
is measured at a fixed wavelength position at 60~\kms\ Doppler offset, while we measure RBE lengths taking the spectral profile into account. This naturally leads to a trend of measuring longer lengths. We confirmed visually that RBE lengths at fixed wavelength filtergrams have a trend of becoming shorter at higher Doppler offset.

Given the striking similarities, we interpret the jet-like features reported by 
\citet{2010ApJ...714L..31G} 
as similar features to what we observe as isolated RBEs scattered over the FOV.

\section{Concluding remarks}

The statistical properties resulting from this dataset, along with better estimated occurrence rates, further strengthen the interpretation of RBEs as the disk-counterpart of type II spicules as first proposed by 
\citet{2008ApJ...679L.167L} 
and further established in Paper~I. 
Given their important role in understanding the outer solar atmosphere, either as seemingly passive tracer for the pervasive presence of Alfv{\'e}nic waves
\citep{2007Sci...318.1574D} 
or more directly as a potential source for mass loading and heating of the corona
\citep{2011Sci...331...55D}, 
further observational study of type II spicules is highly desired in order to advance knowledge of their physical nature. 
An essential property that is not part of this work but can readily be addressed from the analysis method employed on this time series is the spectral evolution of RBEs during their lifetime. 
Such detailed study of the temporal evolution of RBEs is currently under way and will be the subject of a forthcoming paper.  
We further note that observations from the Solar Dynamics Observatory of this target provide the possibility to investigate the response of transition region and coronal diagnostics to RBEs.

\acknowledgments
The Swedish 1-m Solar Telescope is operated by the Institute for Solar
Physics of the Royal Swedish Academy of Sciences in the Spanish
Observatorio del Roque de los Muchachos of the Instituto de
Astrof\'{\i}sica de Canarias. 
This research has made use of NASA's Astrophysical Data System.
B.D.P. was supported through NASA grants NNX08BA99G, NNX08AH45G and NNX11AN98G.

\end{document}